\documentclass[twocolumn]{jpsj2}
\usepackage{graphicx}

\def\runtitle{Dynamical aspects of quantum entanglement}
\def\runauthor{Hiroshi \textsc{Fujisaki}, Atushi \textsc{Tanaka}, 
and Takayuki \textsc{Miyadera}}

\title{%
Dynamical aspects of quantum entanglement
for coupled mapping systems
}

\author{%
Hiroshi \textsc{Fujisaki}\thanks{fujisaki@ims.ac.jp},
Atushi \textsc{Tanaka}$^{1,}$\thanks{tanaka@phys.metro-u.ac.jp},
and 
Takayuki \textsc{Miyadera}$^{2,}$\thanks{Tmdella@aol.com}
}

\inst{%
Department of Theoretical Studies, Institute for Molecular Science,
Myodaiji, Okazaki, 444-8585, Japan \\
$^1$
Department of Physics, Tokyo Metropolitan University,
Minami-Osawa, Hachioji 192-0397, Japan \\
$^2$
Department of Information Sciences, 
Tokyo University of Science, Noda City, 
278-8510, Japan
}

\recdate{\today}

\abst{%
We investigate 
how the dynamical production of 
quantum entanglement for weakly coupled mapping systems 
is influenced by the chaotic dynamics of the corresponding 
classical system.
We derive a general perturbative formula for the entanglement 
production rate which is defined by using the 
linear entropy of the subsystem. 
This formula predicts 
that {\it the increment of the strength of chaos does not 
enhance the production rate of entanglement} 
when the coupling is 
weak enough and the subsystems are strongly chaotic.
The prediction is confirmed by numerical experiments for 
coupled kicked tops and rotors.
We also discuss the entanglement production 
using the Husimi representation of the reduced 
density matrix. 
}

\kword{%
Quantum entanglement, chaos, quantum information processing  
}

\begin{document}
\sloppy
\maketitle

\section{Introduction}

Quantum information processings~\cite{NC00}
have been recognized as a new paradigm of science, 
because of its fundamentality, applicability, and 
interdisciplinarity.
For a quantum computer, a typical example of the paradigm,
to be really effective, it must consist of 
a large number of (interacting) qubits, and its size cannot be microscopic.
Such a ``complex'' structure of an effective 
quantum computer causes many problems due to decoherence 
\cite{decoherence} and ``quantum chaos''.\cite{GS00} 

Here we investigate the relation between 
(quantum) entanglement (which is an essential ingredient of 
quantum information processings) and chaotic behavior of the 
corresponding classical systems. The systems we employ are 
weakly coupled mapping systems, and each subsystem can be 
chaotic in the classical limit.
We can imagine as follows:
A quantum computer has many processing units which 
are ``chaotic'' in some sense (e.g., in the classical limit),
and they weakly interact with each other. 
How much entanglement is generated between the units 
in such a situation? This question is interesting 
from the view point of robustness of quantum computation.\cite{GS00}

Due to the weakness of the coupling, 
we utilize a perturbation theory to analyze the 
entanglement production in the system considered. 
In this paper, we discuss generality of our formula by employing 
two typical chaotic systems, 
i.e., coupled kicked tops and rotors (Sec.~\ref{sec:mapping}). 
Numerical results for both systems are presented 
(Sec.~\ref{sec:numerical}). 
To complement our previous papers,~\cite{TFM02,FMT03} 
we also discuss the wavefunction properties 
of the subsystems in the entanglement production region
employing the Husimi representation (Sec.~\ref{sec:husimi}).

\section{Coupled mapping systems and 
the perturbative formula for entanglement production}
\label{sec:mapping}

Here we consider a composite system which consists 
of two subsystems.
We denote one-step time-evolution operators 
for each subsystem as $U_1$ and $U_2$.
We also introduce another one-step time-evolution operator
$U_{\epsilon}$ which describes the interaction between them. 
Furthermore, the coupling time-evolution 
operator is assumed to be 
\begin{equation}
U_{\epsilon}= \exp \{ -i \epsilon V /\hbar \}
\end{equation}
where $V=\sum_{\alpha} q_{\alpha}^{(1)} \otimes q_{\alpha}^{(2)}$ 
with $q_{\alpha}^{(i)}$ being 
a $\alpha$-th dynamical variable for subsystem $i$,
and $\epsilon$ is a coupling parameter. 
The latter is used in the perturbative treatment below.

Hence the one-step time evolution for the whole system is 
described by
\begin{equation}
  \label{eq:defCKTU}
  |\Psi(t+1) \rangle = 
      U_{\epsilon} U_1 U_2 |\Psi(t) \rangle.
\end{equation}
Since we only examine the case where the whole 
system is in a pure state, 
we quantify the entanglement production by the 
linear entropy of the subsystem 
\begin{eqnarray}
S_{\rm lin}(t) = 1- {\rm Tr}_1 \{ \rho^{(1)}(t)^2 \},
\label{eq:defSlin}
\end{eqnarray}
where $\rho^{(1)}(t)$ is the reduced density
operator for the first subsystem.

Using the time-dependent perturbation theory,
we derive a formula (see Refs.~\citen{TFM02,FMT03} 
for a simpler case and its derivation) 
for the linear entropy: $S_{\rm lin}(t) = 
S^{\rm PT}_{\rm lin}(t) + {\cal O}(\epsilon^3)$ as
\begin{equation}
\label{eq:miyaji}
S^{\rm PT}_{\rm lin}(t) 
= S_0 \sum_{l=1}^t \sum_{m=1}^t D(l,m)
\end{equation}
where $S_0=2 \epsilon^2/\hbar^2$ and 
$D(l,m)$ is a correlation function
of the uncoupled system:
\begin{equation}
  \label{eq:defD}
  D(l,m) = \sum_{\alpha,\beta} 
C^{(1)}_{\alpha,\beta}(l,m) C^{(2)}_{\alpha,\beta}(l,m)
\end{equation}
with
\begin{equation}
\label{eq:defCi}
C^{(i)}_{\alpha,\beta}(l,m) 
= \langle q^{(i)}_{\alpha}(l) q^{(i)}_{\beta}(m) \rangle_i
- \langle q^{(i)}_{\alpha}(l) \rangle_i 
\langle q^{(i)}_{\beta}(m) \rangle_i
\end{equation}
and $q^{(1)}_{\alpha}(l)=(U_1^{l})^{\dagger} q^{(1)}_{\alpha} U_1^{l}$ etc.
represents a free time evolution of the subsystem's variable 
without interaction.
In the next section, 
we apply this result to two model mapping systems, i.e.,
coupled kicked tops and rotors.

\section{Stronger chaos does not imply larger entanglement 
production rate}
\label{sec:numerical}

We introduce two coupled kicked systems as follows.
Coupled kicked tops are described by the 
following unitary operators:
\begin{eqnarray}
U_1 &=& e^{-i k_1 J_{z_1}^2/(2j \hbar)} e^{-i \pi J_{y_1}/(2 \hbar)},
\\
U_2 &=& e^{-i k_2 J_{z_2}^2/(2j \hbar)} e^{-i \pi J_{y_2}/(2 \hbar)},
\\
U_{\epsilon} &=& e^{-i \epsilon J_{z_1} J_{z_2}/(j \hbar)},
\end{eqnarray}
and coupled kicked rotors are by the following:
\begin{eqnarray}
U_1 &=& e^{-i I_1^2/(2 \hbar)} e^{-i k_1 \cos \theta_1/\hbar},
\\
U_2 &=& e^{-i I_2^2/(2 \hbar)} e^{-i k_2 \cos \theta_2/\hbar},
\\
U_{\epsilon} &=& e^{-i \epsilon \cos(\theta_1-\theta_2)/\hbar},
\end{eqnarray}
where $k_1$, $k_2$ represent the strength of nonlinearity
related to chaotic properties of the systems.
For the detailed information of the former and latter systems, 
see Refs.~\citen{TFM02,FMT03} and Ref.~\citen{TAI89}, respectively.
We note that the periodic boundary
conditions with period $2 \pi$ 
are imposed for both variables $\theta_i$ and $I_i$.

\begin{figure}[tb]
\begin{center}
\includegraphics[scale=0.8]{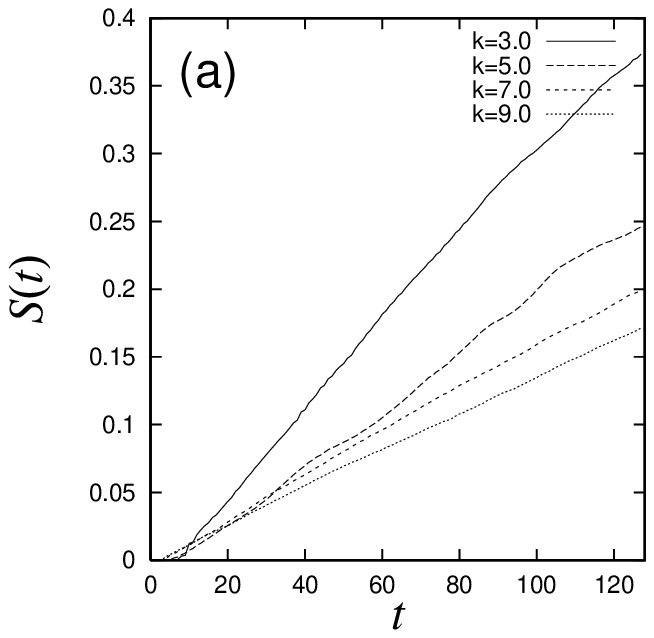}
\includegraphics[scale=0.8]{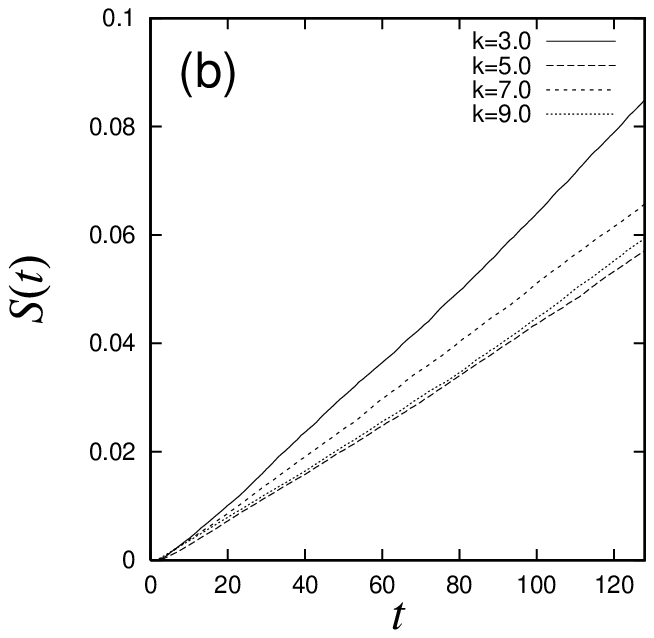}
\caption{\label{fig:linear}
Time evolution of the linear entropy 
for the coupled kicked tops (a) and rotors (b).
The initial states are embedded in chaotic seas in both 
systems. For simplicity, we take $k_1=k_2=k$.
The interaction strength is weak 
($\epsilon=10^{-4}$ for tops and $\epsilon=10^{-3}$ for rotors).
$\hbar=1.0$ and $j=80$ for tops and $\hbar= 2\pi/128$ for rotors.
}
\end{center}
\end{figure}

As an initial state,  
we take a product, i.e., separable state 
of spin coherent states\cite{FMT03}
(coherent states)
for coupled kicked tops (rotors).
In Fig.~\ref{fig:linear}, 
according to Eq.~(\ref{eq:defSlin}), we calculate the time 
evolution of the linear entropy 
for the ``chaotic'' initial conditions, 
i.e., their corresponding 
classical states in phase space are embedded in chaotic seas.
For {\it both} cases,
the linear entropy increases linearly as a function of time 
in this transient region, which is called 
$t$-linear entanglement production region in the following, 
before the ``equilibration'' of the entropy. 
In these cases, the coupling is in a sense weak, and 
we apply the formula, Eq.~(\ref{eq:miyaji}), 
to these situations.

\begin{figure}[tb]
\begin{center}
\includegraphics[scale=0.8]{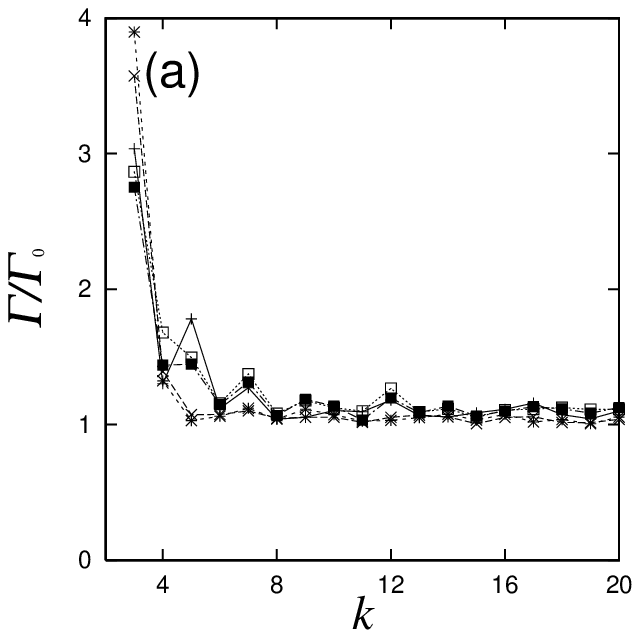}
\includegraphics[scale=0.8]{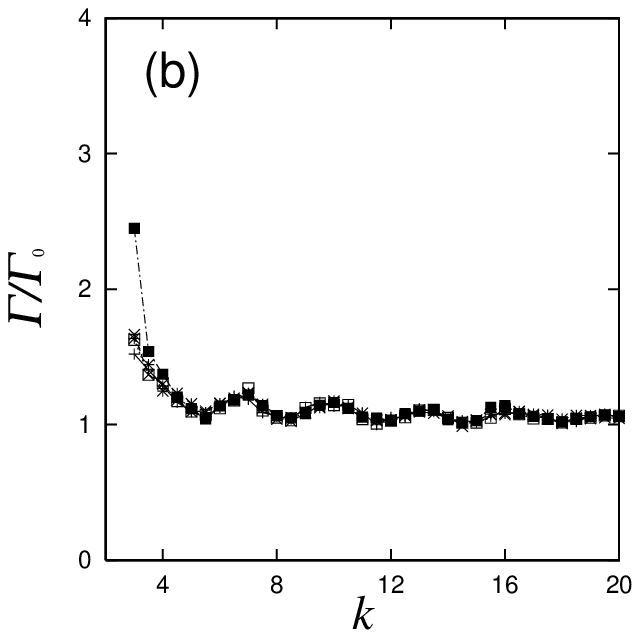}
\caption{\label{fig:nonlinear}
Nonlinear parameter dependence of the normalized 
entanglement production rate ($\Gamma/\Gamma_0$) 
for tops (a) and rotors (b).
The plots for five different initial conditions 
embedded in the chaotic seas are superimposed.
We take $k_1=k_2=k$.
The $k$ dependency of $\Gamma$ for the coupled rotors 
exhibits oscillations. We suggest that the 
oscillations are similar ones that are observed in the 
diffusion constant of a kicked rotor 
due to higher-order time correlations\cite{Rechester80}.
}
\end{center}
\end{figure}

From a classical intuition, we expect that 
the correlation functions $D(l,m)$ decay very quickly 
as a function of $|l-m|$
for strongly chaotic cases (this is numerically confirmed for 
a kicked top\cite{TFM02,FMT03} and a kicked rotor\cite{Shepelyansky83}), 
so it is assumed to be 
\begin{equation}
  \label{eq:Dpheno}
  D(l,m) \simeq  D_0 e^{-\gamma|l-m|}. 
\end{equation}
Using above, we can easily derive the following 
for the entanglement production rate:
\begin{equation}
\label{eq:dMdt}
\Gamma \equiv
\left. \frac{d S^{\rm PT}_{\rm lin}(t)}{dt} \right|_{t \gg 1/\gamma} 
\simeq
\Gamma_0 \coth (\gamma/2)
\end{equation}
with $\Gamma_0=S_0 D_0$.
Since $\gamma$ often becomes large as the nonlinear parameters increase,  
and there is a numerical experiment 
that  $\gamma$ and the sum 
of positive Lyapunov exponents are correlated in the 
coupled kicked tops~\cite{FMT03},
this formula implies that strong chaos ($\gamma \rightarrow \infty$) 
leads to a saturation of the entanglement production rate 
($\Gamma \rightarrow \Gamma_0$).   
This prediction is actually confirmed by numerical 
experiments for {\it both} systems 
(coupled kicked tops and rotors) 
as shown in Fig.~\ref{fig:nonlinear}.

The above is a result for strongly chaotic cases.  
In contrast, we comment for weakly chaotic cases using 
the above formula: 
For strongly chaotic cases {\it with bounded phase space}, 
$D_0$ saturates to a certain value determined by a 
statistical argument, whereas, for weakly chaotic cases,
$D_0$ grows up as the nonlinear parameters increase.
This roughly explains the common belief:
{\it chaos enhances entanglement for weakly chaotic systems}. 
(See the cited references in Refs.~\citen{TFM02,FMT03}.)
This seems to contradict our result, 
but the point is that we focus on strongly chaotic 
cases where $D_0$ saturates, and in such a situation, 
stronger chaos saturates the entanglement production rate 
for weakly coupled mapping systems.
We also comment that this formula can be easily extended 
to describe flow systems with continuous time~\cite{FMT03}, 
which is useful when applying to more realistic systems.

\section{Husimi representation as 
a tool to study entanglement production}
\label{sec:husimi}

The linear entropy which we employ as a measure of 
entanglement is calculated from the reduced 
density operator $\rho^{(1)}(t)$.
Hence it is natural to ask what happens in  
$\rho^{(1)}(t)$ itself when the entanglement 
production occurs.
Here we shall address this issue 
using the results of coupled kicked tops.

\begin{figure}[tb]
\begin{center}
\includegraphics[scale=1.0]{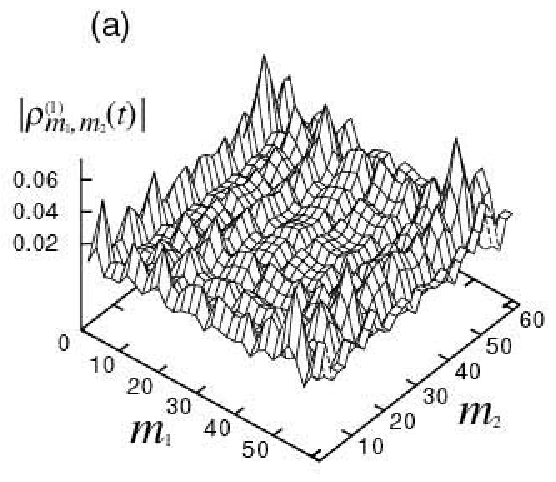}
\includegraphics[scale=1.0]{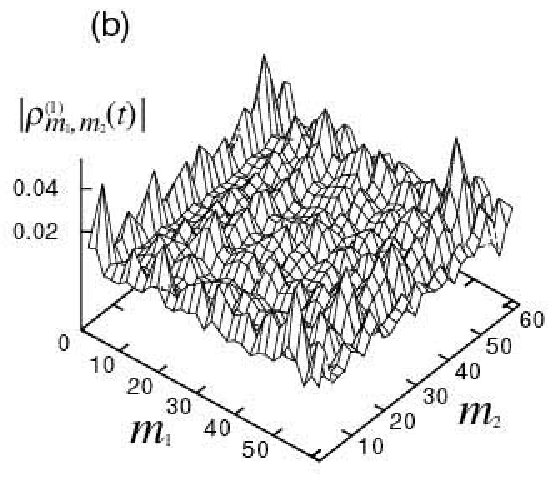}
\caption{\label{fig:wave}
The contour plot of the absolute value of  
(a) the density matrix at $t=112$ for a kicked top, and
(b) the reduced density matrix $t=112$ for coupled kicked tops
with $\epsilon=0.003$. 
The initial state is embedded in the chaotic sea, and 
given by $|\theta,\phi \rangle= |0.89,0.63 \rangle$ 
for (a) and $|\theta_1,\phi_1 \rangle \otimes |\theta_2,\phi_2 \rangle 
= |0.89,0.63 \rangle \otimes |0.89,0.63 \rangle$ for (b).
}
\end{center}
\end{figure}

In Fig.~\ref{fig:wave} (b),
we show the absolute value 
of the reduced density matrix 
$\rho^{(1)}_{m_1,m_2}(t) 
=
\langle j m_1 | \rho^{(1)}(t) | j m_2 \rangle$
during the $t$-linear entanglement production region
(see Sec.~\ref{sec:numerical}).
We also show the corresponding density matrix of a single kicked 
top [Fig.~\ref{fig:wave} (a)].
Here $|jm \rangle$ is the simultaneous eigenstate 
of $J^2$ and $J_z$ for top 1.
From the previous works~\cite{SKO96}, 
we expect that the decay of the off-diagonal elements
of the reduced density matrix reflects entanglement production.
Such an analysis is useful when the entanglement production 
becomes almost equilibrated,
however, two cases in Fig.~\ref{fig:wave} are very hard to 
distinguish since 
the system is in  
the $t$-linear entanglement production region.
(Of course, after a long time, 
the off-diagonal elements finally decay.)
To solve this issue, 
we propose to examine the Husimi representation of the 
density operator~\cite{Takahashi}.

\begin{figure}
\hfill
\begin{center}
\includegraphics[scale=0.8]{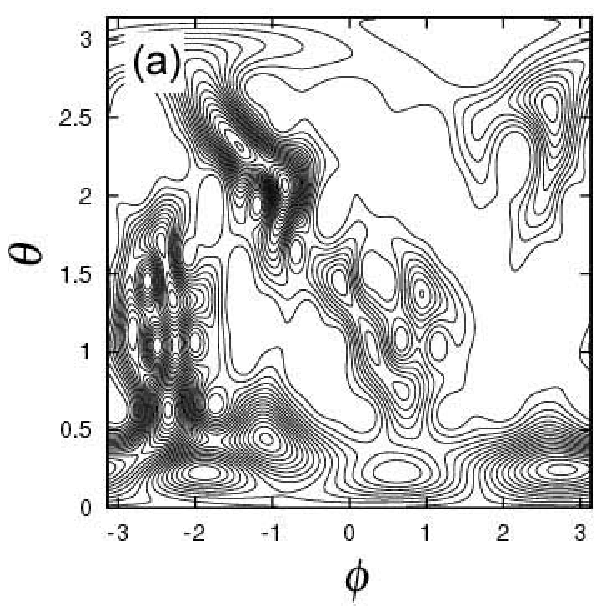}
\includegraphics[scale=0.8]{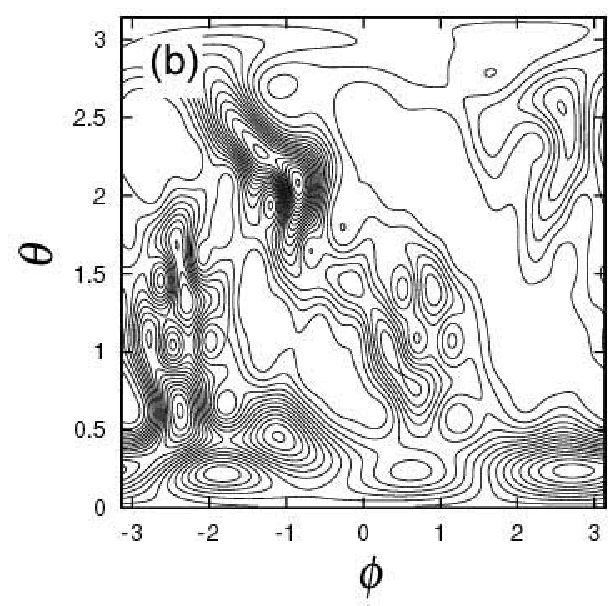}
\caption{\label{fig:husimi1}
The contour plots of 
the Husimi functions corresponding to 
the density matrices in Fig.~\ref{fig:wave}. 
The scale is normal.
}
\end{center}
\end{figure}

\begin{figure}
\hfill
\begin{center}
\includegraphics[scale=0.8]{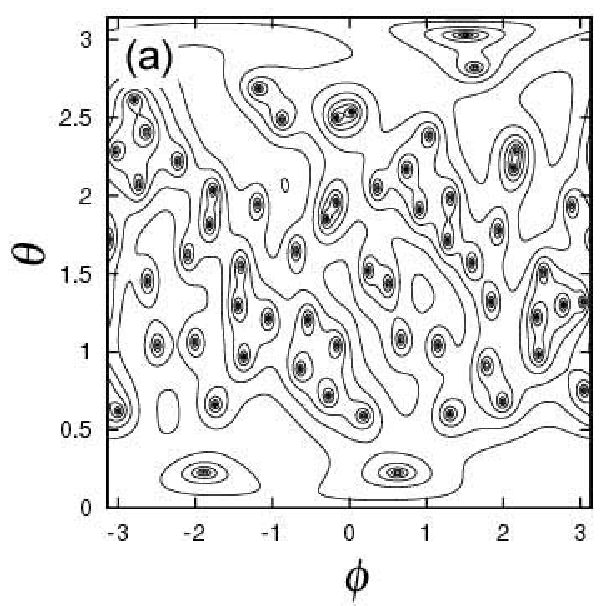}
\includegraphics[scale=0.8]{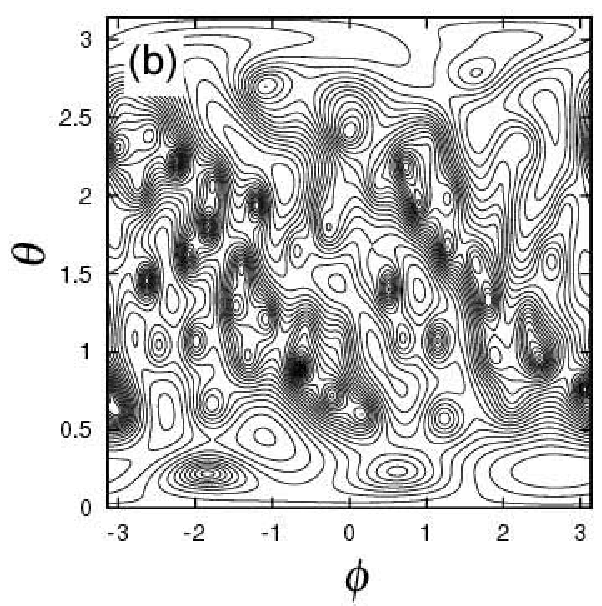}
\caption{\label{fig:husimi2}
The contour plots of 
the Husimi functions 
corresponding to the density matrices in Fig.~\ref{fig:wave}. 
The scale is logarithmic.
Note that the number of the Husimi zeros 
is $2 j=60$ [see (a)].
}
\end{center}
\end{figure}

The Husimi function for the kicked top system is defined by 
\begin{equation}
H(\theta,\phi)= 
\langle \theta,\phi | \rho^{(1)}(t) | \theta,\phi \rangle,
\end{equation}
where $|\theta,\phi \rangle$ is a spin-coherent state given by
\begin{equation}
\langle j m| \theta,\phi\rangle
=\frac{\gamma^{j-m}}{(1+|\gamma|^2)^j} \sqrt{\frac{2j!}{(j+m)!(j-m)!}}
\end{equation}
with $\gamma=e^{i \phi} \tan (\theta/2)$.
For simplicity of visualization, 
we take a smaller value of $j$ ($j=30$ instead of $j=80$ used in 
Figs.~\ref{fig:linear} and \ref{fig:nonlinear}).
Though the normal-scale plots do not show 
any significant fingerprint of entanglement production 
due to the interaction (Fig.~\ref{fig:husimi1}), 
the log-scale plots do show it (Fig.~\ref{fig:husimi2}).
The Husimi zeros are easily confirmed in Fig.~\ref{fig:husimi2} (a).
They become local minima with positive values 
as the interaction strength gets larger
[Fig.~\ref{fig:husimi2} (b)].
Husimi zeros are considered to characterize ``complexity''
of the wavefunction~\cite{LVT}, 
so the quantification of this local minima, 
the traces of the Husimi zeros,
might characterize the entanglement production 
from ``complexity theory'' point of view~\cite{Adachi}.
Many measures for complexity of a Husimi function 
have been proposed so far~\cite{LVT,Takahashi}, but 
we suppose that another measure, e.g., 
the phase-space averaged curvature of a Husimi function,
is needed to quantify the complexity related to entanglement production.
This issue will be discussed elsewhere.


\section{Summary and outlook}
\label{sec:summary}

We have investigated how the production rate of 
quantum entanglement  
is affected by the chaotic properties of the corresponding 
classical system using coupled kicked tops and rotors.
We have derived and numerically confirmed that  
{\it the increment of the strength of chaos does not 
enhance the production rate of entanglement} 
when the coupling is weak enough and the subsystems are strongly chaotic.
We believe that this conclusion is general and applicable 
to any mapping system {\it with bounded phase space}.
It is often believed that {\it chaos enhances entanglement},
but that can be the case for weakly chaotic regions, 
and such effects can be also explained by the 
factor $D_0$ in our formula, Eq.~(\ref{eq:dMdt}). 
Note again that the above our conclusion is basically for 
strongly chaotic regions where $D_0$ saturates.
We also proposed to use the logarithmic plot of the 
Husimi representation of a reduced density operator 
to investigate entanglement production in coupled 
quantum systems.

Since the dynamical aspects of entanglement production 
have been studied in relation with quantum computing\cite{BS03},
we hope that our analysis will be useful in the studies of 
quantum information processsings.
In Ref.~\citen{GS00}, Prosen and Znidaric showed 
that quantum computing can be more robust 
with the use of quantum chaotic systems 
compared to the corresponding regular systems.
We expect that the saturation of entanglement production 
found here will be utilized in such a situation.
In addition, entanglement production is strongly related 
to decoherence processes \cite{decoherence}, 
so we expect that there are many applications of 
our results for quantum dynamical processes 
like quantum optical processes in environments\cite{MS98} 
or chemical reactions in solvents\cite{Okazaki01}. 

One of the authors (H.F.) thanks 
T.~Prosen, K.~Takahashi, M.~Toda, S.~Kawabata, and K.~Saito for 
useful comments.


\begin{thebibliography}{99}

\bibitem{NC00}
M.~A.~Nielsen and I.~L.~Chuang,
{\it Quantum Computation and Quantum Information}
(Cambridge University Press, Cambridge, 2000);
G.~P.~Berman, G.~D.~Doolen, R.~Mainieri, and 
V.~I.~Tsifrinovich,
{\it Introduction to Quantum Computers} 
(World Scientific, Singapore, 1998).

\bibitem{decoherence}
D.~Giulini, E.~Joos, C.~Keifer, J.~Kupsch, I.-O.~Stamatescu,
and H.D.~Zeh, {\it Decoherence and the Appearance of a Classical World
in Quantum Theory} (Springer, Berlin, 1996);
W.~H.~Zurek, 
Physics Today, {\bf 44} (1991) 36;
Prog.~Theor.~Phys. {\bf 89} (1993) 281;
e-print quant-ph/0105127.

\bibitem{GS00}
B.~Georgeot and D.~L.~Shepelyansky,
Phys.~Rev.~E {\bf 62} (2000) 3504;
T.~Prosen and M.~Znidaric, 
J.~Phys.~A: Math.~Gen.~ {\bf 34} (2001) L681, 
and references cited therein. 

\bibitem{TFM02}
  A.~Tanaka, H.~Fujisaki, and T.~Miyadera,
  Phys.~Rev.~E {\bf 66} (2002) 045201(R).


\bibitem{FMT03}
  H.~Fujisaki, T.~Miyadera, and A.~Tanaka,
  Phys.~Rev.~E (submitted); e-print quant-ph/0211110. 


\bibitem{TAI89}
  M.~Toda, S.~Adachi, and K.~Ikeda,
  Prog.~Theor.~Phys.~Suppl. {\bf 98} (1989) 323.

\bibitem{Shepelyansky83}
D.~L.~Shepelyansky, Physica D {\bf 8} (1983) 208.

\bibitem{Rechester80}
A.~B.~Rechester and R.~B.~White,
Phys.~Rev.~Lett.~ {\bf 44} (1980) 1586; 
ibid. {\bf 45} (1980) 851 (Erratum).

\bibitem{LVT}
P.~Leboeuf and A.~Voros, J.~Phys.~A {\bf 32} (1990) 1765;
M.~Toda, Physica D {\bf 59} (1992) 121.

\bibitem{Adachi}
S.~Adachi, private communication.

\bibitem{SKO96}
M.~Sakagami, H.~Kubotani, and T.~Okamura,
Prog.~Theor.~Phys. {\bf 95} (1996) 703;
A.~Lakshminarayan,
Phys.~Rev.~E {\bf 64} (2001) 036207.

\bibitem{Takahashi}
K.~Takahashi, 
  Prog.~Theor.~Phys.~Suppl. {\bf 98} (1989) 109.

\bibitem{BS03}
S.~Bettelli and D.~L.~Shepelyansky,
e-print quant-ph/0301086.


\bibitem{MS98}
P.~Meystre and M.~Sargent III,
{\it Elements of Quantum Optics}, 3rd ed. 
(Springer-Verlag, Berlin, 1998).

\bibitem{Okazaki01}
S.~Okazaki, Adv.~Chem.~Phys.~{\bf 118} (2001) 191.

\end{thebibliography}
\end{document}